\documentclass[12pt,preprint]{aastex}

\newcommand{\beq}{\begin{equation}}
\newcommand{\eeq}{\end{equation}}

\def\3he{$^3$He\,}
\def\he4{$^4$He\,}

\shorttitle{Electron Acceleration}
\shortauthors{Sharykin \& Liu}

\begin{document}

\title{Onset of Electron Acceleration in a Flare Loop}

\author{
Ivan Sharykin\altaffilmark{1,2},
Siming Liu\altaffilmark{1}
, and Lyndsay Fletcher\altaffilmark{3}
}

\altaffiltext{1}{Key Laboratory of Dark Matter and Space Astronomy, Purple Mountain Observatory, Chinese Academy of Sciences, Nanjing, 210008, China, liusm@pmo.ac.cn}
\altaffiltext{2}{Space Research Institute (IKI) of Russian Academy of Science, Profsoyuznaya Str. 84/32, 117997, Moscow, Russia}
\altaffiltext{3}{School of Physics and Astronomy, SUPA, University of
Glasgow, Glasgow, G12 8QQ, UK}

\begin{abstract}

We carried out detailed analysis of X-ray and  radio observations of a simple flare loop that occurred on 12th August 2002, with the impulsive hard X-ray (HXR) light curves dominated by a single pulse. The emission spectra of the early impulsive phase are consistent with an isothermal model  in the coronal loop with a temperature reaching several keVs. A power-law high-energy spectral tail is evident near the HXR peak time, in accordance with the appearance of  footpoints at high energies, and is well correlated with the radio emission. The energy content of the thermal component keeps increasing gradually after the disappearance of this nonthermal component. These results suggest that electron acceleration only covers a central period of a longer and more gradual energy dissipation process and that the electron transport within the loop plays a crucial role in the formation of the inferred power-law electron distribution. The spectral index of power-law photons shows a very gradual evolution indicating a quasi-steady state of the electron accelerator, which is confirmed by radio observations. These results are consistent with the theory of stochastic electron acceleration from a thermal background. Advanced modeling with coupled electron acceleration and spatial transport processes is needed to explain these observations more quantitatively, which may reveal the dependence of the electron acceleration on the spatial structure of the acceleration region.

\end{abstract}


\section{INTRODUCTION}
\label{intro}

Acceleration of particles is one of the key issues in the study of solar flares \citep{2013arXiv1307.1837C}. Although it is well established that charged particles need to be accelerated via electro-magnetic interactions and energetic neutrals are produced by energetic charged particles via secondary processes \citep{2009ApJ...693L..11M}, the exact nature of particle acceleration is still a matter of debate \citep[e.g.,][]{2008ApJ...675.1645F, 2009A&A...508..993B, 2011SSRv..159..357Z}. Given the complexity of the magnetic field structure in the solar atmosphere, where most of the particles are presumably accelerated, current observations suggest that the exact process of particle acceleration may be coupled with the evolution of the magnetic field so that a variety of particle acceleration processes operate in flares and different processes play the dominant role in different stages and/or parts of solar flares \citep{2013ApJ...769..135L}. For example, for the classical large two-ribbon flares, energetic charged particles may be accelerated by large scale electric fields in the reconnecting current sheet \citep{1994ApJ...435..469B,2004ApJ...604..884Z}, by a newly formed collapsing loop via the betatron process \citep{1997ApJ...485..859S, 2012A&A...546A..85G}, by Alfv\'{e}n waves within the loop \citep{2008ApJ...675.1645F}, by shocks induced by the reconnection outflows \citep{1997ApJ...485..859S, 2006A&A...454..969M, 2013PhRvL.110e1101N}, and by turbulent plasma waves generic to the energy dissipation process \citep{2004ApJ...610..550P}. Comprehensive observational coverage is therefore necessary to understand the processes of particle acceleration in detail, especially for large flares with complex structures.

Observational study of particle acceleration may be simplified dramatically by focusing on flares with a simple geometrical structure, implying a relatively simple magnetic field, in particular for individual flare loops. Based on {\it RHESSI} observations of dense flare loops, \citet{2012A&A...543A..53G} recently showed that the length of the acceleration region in these events is about half the length of the flare loops and the rate of acceleration defined as particles per second can be very high \citep{2013ApJ...766...28G}. Such flares may be triggered by the internal kink instability of a magnetic loop  \citep{2007A&A...467..327H}, and comprehensive observations and analysis of these flares play an important role in quantitative modeling of flares and detailed studies of the related physical processes \citep{1984ApJ...279..882E, 2008ApJ...682.1351K, 2009ApJ...702.1553L, 2009A&A...494.1127R, 2011A&A...525A..57P}.

In a recent paper \citep{2013ApJ...769..135L}, we showed that a {\it GOES} class B flare loop produced super-hot plasmas with an impulsive temperature evolution with little evidence of particle acceleration \citep{1979ApJ...228..592B}. In the context of stochastic acceleration of particles by plasma waves from the thermal background plasma \citep{2004ApJ...610..550P, 2012SSRv..173..535P}, \citet{2010ApJ...709...58L} proposed a simple one-zone model, where the spatially integrated particle distribution is modeled with the detailed structure of the acceleration region ignored, for the so-called elementary energy release event to account for the frequently observed soft-hard-soft spectral evolution of HXR bursts \citep{2004A&A...426.1093G, 2006ApJ...651..553Z, 2011SSRv..159..107H}. The model may be applied to simple flare loops as well. \citet{2004ApJ...610..550P} showed that the acceleration of particles is very sensitive to the intensity of the turbulent plasma waves so that a purely thermal X-ray flare may suggest that the turbulent intensity is not strong enough to produce sufficient energetic electrons to be detected by {\it RHESSI}. Nevertheless the complex source structure along the loop in the flare analyzed by \citet{2013ApJ...769..135L} does indicate that this simple model may not capture the physical process responsible for the impulsive temperature evolution, and that spatial structure needs to be considered in detailed modeling of specific flares \citep{2009ApJ...702.1553L}.

In this paper, we analyze X-ray observations with {\it GOES} and {\it RHESSI}, and radio observations with Nobeyama radio polarimeters (NoRP), of a {\it GOES} class C1.4 flare loop with clear hard X-ray (HXR) emission above 50 keV implying efficient electron acceleration. The flare occurred on 12th August 2002 and has been studied previously by \citet{2009RAA.....9.1155L}, who focused on the chromospheric evaporation process. We focus on the energetics, the characteristics of emissions produced by nonthermal electrons and the relation between the thermal and the nonthermal component to explore the overall energy release process. Our observational results are given in Section \ref{observation}. In Section \ref{dis}, we discuss the implications of these results on the energy release and electron acceleration processes. The conclusion is drawn in Section \ref{con}.


\section{Observations}
\label{observation}

\begin{figure}[htb]
\begin{center}
\includegraphics[width=10cm]{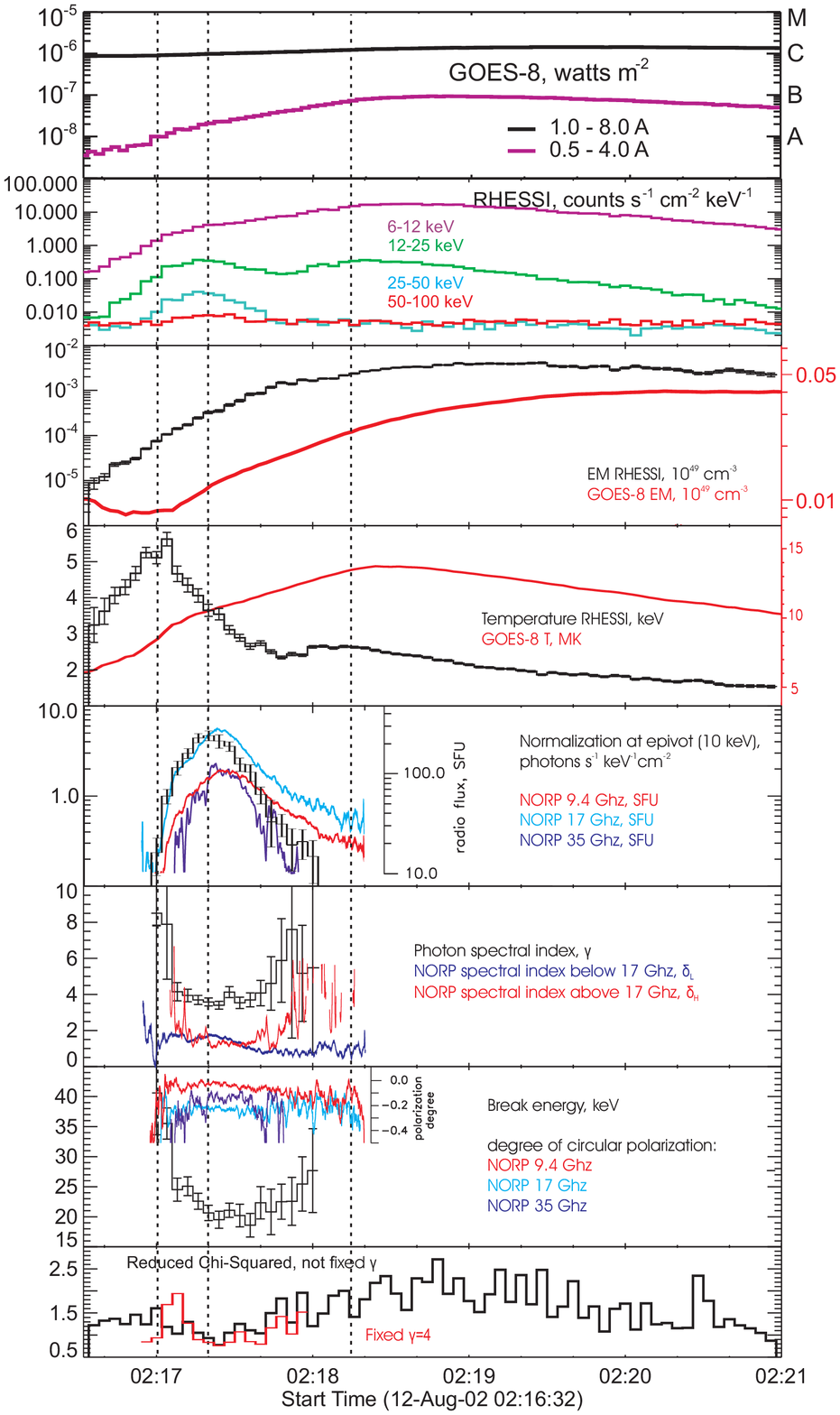}
\caption{
Observational summary. The 1st and 2nd panels give the {\it GOES} and {\it RHESSI}
 light curves, respectively. The 3rd and 4th panels are for the {\it RHESSI} (black) and {\it GOES} (blue) $EM$ and temperature, respectively.
 The 5th panel gives the normalization of the broken power-law component at the pivot energy of 10 keV and the radio fluxes. The 6th panel shows the spectral indexes.
 The 7th panel gives the break energy and the polarization fraction of the radio emission. The bottom panel gives the reduced $\chi^2$ of the {\it RHESSI} spectral fit. The red segment is for a model with the spectral index above the break energy fixed at 4.0.
}
\label{f1}
\end{center}
\end{figure}

The flare occurred on 12th August 2002 and had a preflare {\it GOES} background fluxes of $8.5\times 10^{-7}$ and $9.5\times 10^{-9}$ W m$^{-2}$ in the 1-8 \AA\ (1.5-12 keV) and 0.5-4 \AA\ (3-24 keV) channel, respectively. The {\it GOES} high energy channel flux started to increase at 02:16:30 UT while the low energy channel flux started to increase about half a minute later at 02:17:00 UT (First panel of Fig. \ref{f1}). \citet{2012ApJS..202...11R} argued that the background fluxes subtracted from the {\it GOES} fluxes for spectral analysis should be chosen to guarantee a smooth increase of the derived temperature and emission measure ($EM$) in the early rise phase. We find a low and a high channel background flux of $7.5\times 10^{-7}$ and $5.6 \times 10^{-9}$ W m$^{-2}$, respectively, and derive the {\it GOES} temperature (3rd panel of Fig. \ref{f1}) and $EM$ (4th panel of Fig \ref{f1}.) using the standard IDL GOES analysis software.
Due to the high preflare background, results before 02:17:00 UT are questionable.

 The {\it RHESSI} count rates in the 6-12 keV and 12-25 keV band start to increase before 02:16:30UT (Second panel of Fig. \ref{f1}). Whenever acceptable, we use an isothermal model in the OSPEX software to fit the {\it RHESSI} photon spectra above 6 keV  for every 4 second interval obtained with the front segment of detectors 1, 3, and 6 \citep{2011ApJ...728....4G, 2013ApJ...769..135L}. A broken power-law component is added to the fitting model when the reduced $\chi^2$ of the isothermal model exceeds 3. For this broken power-law component \citep{2005A&A...435..743S}, the photon spectral index below the break energy is fixed at 1.5 with $A$: the normalization at a pivot energy of 10 keV, $\epsilon_b$: the break energy, and $\gamma$: the power-law index above the break energy as free parameters. The derived $EM$, temperature, $A$, $\gamma$, and $\epsilon_b$ are shown in  panels 3 to 7 of Figure \ref{f1}, respectively. The last panel shows the reduced $\chi^2$. We also study the effect of pileup correction. Although the pileup correction can improve the $\chi^2$ fitting, the model parameters do not change significantly.

Panels 5 to 7 also show the radio flux measurement at 9.4, 17, and 35 GHz,  spectral indices derived from the relevant flux ratios, and polarization fraction, respectively. The radio polarization fraction and the good correlation between the radio fluxes and $A$ show clearly that the radio emission is produced by energetic electrons via the gyrosynchrotron process.
The gyrosynchrotron radio emission is mostly produced by electrons well above 50 keV, which may explain the coincidence of peak times of the $50-100$ keV and radio fluxes. The delay of these peaks with respect to the HXR peak at lower energies may be attributed to more efficient trapping of higher energy electrons in the coronal loop \citep{2002ApJ...572..609L,0004-637X-673-1-598}, which may also explain the fact that the 9.4 and 17 GHz fluxes decay more gradually than $A$ \citep{2013ApJ...768..190F}. Alternatively, the high energy tail of the thermal component may contribute significantly to the low frequency radio emission via the gyrosynchrotron process in the decay phase, which is consistent with the faster decay of 35 GHz flux than those of the other two low frequency channels \footnote{The radio spectral shape between 17 and 35 GHz is not consistent with a bremsstrahlung origin of the emission.}.
 More detailed modeling is needed to clarify this issue \citep{2000ApJ...533.1053N}.
The first vertical dashed line marks when the broken power-law component is first introduced and radio fluxes start to increase dramatically. We associate this with the onset of electron acceleration. The second dashed line marks the peak of $A$. The acceleration phase is identified with the period when the broken power-law component is needed. As we will see below, this high-energy component is associated with the footpoints.

\begin{figure}[ht]
\begin{center}
\includegraphics[width=12cm]{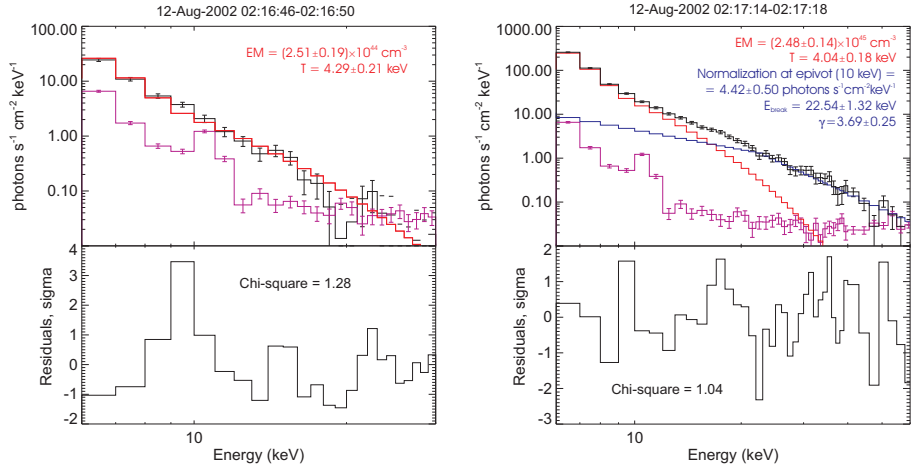}
\caption{
Photon (upper) spectra of a 4-second interval in the preheating (left) and acceleration (right) phases. The red, blue, magenta, and black lines indicate the thermal component, broken power-law component, pre-flare background, and background-subtracted data, respectively. Model parameters are indicated on the figures. The lower panels show the normalized residuals.
}
\label{f2}
\end{center}
\end{figure}

\begin{figure}[ht]
\begin{center}
\includegraphics[width=11cm]{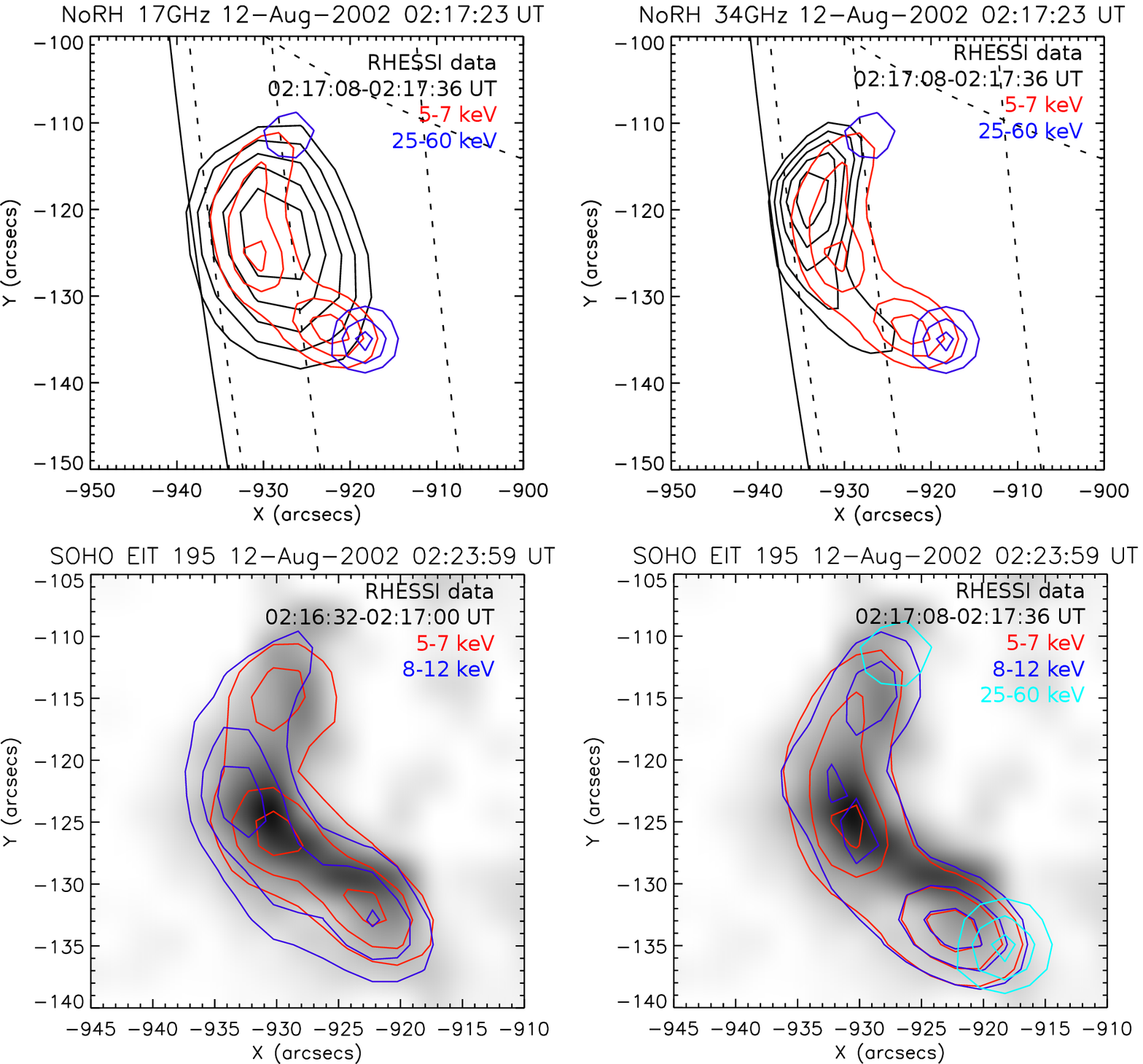}
\caption{
Upper: Radio and X-ray image contours during the acceleration phase. The radio contours are 50, 60, 70, 80, and 90\% of the peak brightness constructed with the CLEAN algorithm.
Lower: EIT image in the decay phase showing the post flare loop structure and {\it RHESSI} image contours in the preheating phase (left) and acceleration phase (right). The X-ray contours are 50, 70, and 90\% of the peak value constructed with the CLEAN algorithm for the front segment of detectors 1, 3, 4, 5, 6, 7, 8, and 9.
}
\label{f3}
\end{center}
\end{figure}

Details of the {\it RHESSI} photon spectral fit at two 4 second time intervals are shown in Figure \ref{f2}. The left panel corresponds to an interval in the preheating phase when one thermal component can give a reasonable fit to the observed photon spectrum. The right panel corresponds to the acceleration phase when the high energy emission is dominated by a broken power-law component. However emission below 10 keV is always dominated by the isothermal component, similar to a flare studied by \citet{2011ApJ...728....4G}. The reduced $\chi^2$ of both fits are less than 1.5 and the corresponding model parameters are indicated in the Figure.

 The upper panels of Figure \ref{f3} show the radio image contours over-plotted with X-ray contours in the acceleration phase. The X-ray emission clearly outlines a loop structure. Most of the radio emission comes from the northern leg of the loop, which is also the location of the weaker footpoint HXR emission. Both of these are consistent with a stronger magnetic field at this footpoint that inhibits electron precipitation in the `trap-plus-precipitation' scenario \citep{1976MNRAS.176...15M, 2002ApJ...572..609L}.  The lower panels show the {\it RHESSI} image contours in different energy bands  over-plotted on an image of the post impulsive phase loop obtained with the EIT at 195 \AA\ at 02:23:59 UT. The appearance of footpoints at high energies ($>25$ keV) in the acceleration phase does not affect the structure of the thermal coronal source significantly and the {\it RHESSI} and {\it GOES} thermal components appear to occupy the same volume as outlined by the contours in the 5-7 keV and 8-12 keV bands.
There were no accompanying CMEs and type II radio bursts as seen in eruptive events and no type III radio bursts, in agreement with the simple magnetic topology of a closed loop.

\section{Results and Discussion}
\label{dis}

In this section, we explore the energetics and electron acceleration process in this flare loop based on the observational results presented above. To measure the relevant physical quantities, we adopt the flare loop model explored numerically by \citet{2009ApJ...702.1553L}, where the loop is separated into three distinct regions: the loop top acceleration site, the legs of the loop where the transport processes dominate, and the dense chromospheric footpoints.

\subsection{Energetics}

Panels 3 and 4 of Figure \ref{f1} show that the $EM$ of the {\it RHESSI} thermal component is at least one order of magnitude lower than that of the {\it GOES} and the temperature of the {\it RHESSI} thermal component is at least 70\% higher than the {\it GOES} temperature. Given the gradual evolution of these parameters and the relatively stable structure of the soft X-ray source, one may associate the {\it RHESSI} and {\it GOES} thermal components with the loop top acceleration site and the legs of the loop, respectively and assume that the {\it RHESSI} and {\it GOES} thermal components are in pressure equilibrium \citep{1978ApJ...220.1137A, 2009ApJ...702.1553L}. Then from pressure balance we have $$T_{\rm RHESSI} (EM_{\rm RHESSI}/V_{\rm RHESSI})^{1/2} = T_{\rm GOES} (EM_{\rm GOES}/V_{\rm GOES})^{1/2},$$ where $V_{\rm RHESSI}$ and $V_{\rm GOES}$ are the volume of the {\it RHESSI} and {\it GOES} thermal plasmas, respectively.

\begin{figure}[ht]
\begin{center}
\includegraphics[width=11cm]{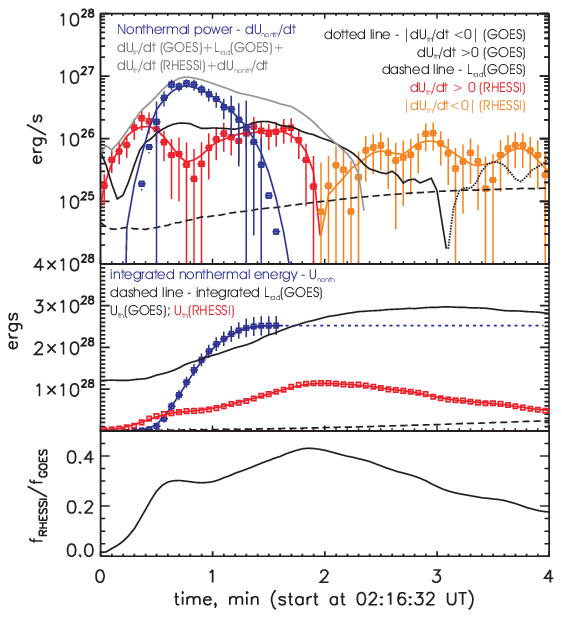}
\caption{
Top: time profiles of different power components. Here we have assumed a constant total volume $V_t$ of $10^{26}$ cm$^{3}$ for the {\it GOES} and {\it RHESSI} thermal components. The nonthermal power is for electrons above $\epsilon_b$. $L_{\rm rad}(\rm GOES)$ indicates the luminosity of the {\it GOES} thermal component. The lines are running smooth of the corresponding data points.
Middle: time profiles of different energy components. The dashed line is for the time integrated {\it GOES} luminosity. The blue data points are for the integrated energy deposition rate of nonthermal electrons. Bottom: time profile of the ratio of the volume filling factor of the {\it RHESSI} and {\it GOES} thermal components.
}
\label{f4}
\end{center}
\end{figure}

The X-ray images in Figure \ref{f3} show that the volume of the soft X-ray emitting plasmas is on the order of $10^{26}$ cm$^{-3}$ assuming a line-of-sight thickness comparable to the width of the loop. Adopting a fiducial total volume $$V_t= V_{\rm RHESSI} + V_{\rm GOES} \sim 10^{26} {\rm cm}^{-3}$$ for these thermal components, Figure \ref{f4} shows the energetics of different components observed in X-rays and the evolution of the ratio of the volume filling factors $$f_{\rm RHESSI}=V_{\rm RHESSI}/V_t$$ and $$f_{\rm GOES}=V_{\rm GOES}/V_t$$ for the {\it RHESSI} and {\it GOES} thermal plasmas, respectively, where the thermal energy $$U_i = 3 k_{\rm B}T_i(EM_i V_t f_i)^{1/2}$$  with $i$ representing {\it RHESSI} and {\it GOES} and $k_{\rm B}$ the Boltzmann constant. For the sake of simplicity,  we also assume that $$f_{\rm RHESSI}+f_{\rm GOES}=1,$$ which is appropriate given the relatively stable structure of the flare loop \citep{2009RAA.....9.1155L} and the dramatic difference between the {\it RHESSI} and the {\it GOES} $EM$. Then we have $$f_{\rm RHESSI} = T_{\rm RHESSI}^2 EM_{\rm RHESSI}/(T_{\rm RHESSI}^2 EM_{\rm RHESSI}+T_{\rm GOES}^2 EM_{\rm GOES})$$ and $$f_{\rm GOES} = T_{\rm GOES}^2 EM_{\rm GOES}/(T_{\rm RHESSI}^2 EM_{\rm RHESSI}+T_{\rm GOES}^2 EM_{\rm GOES}).$$

Using the classical thick-target HXR emission model, one can derive the energy deposition rate by energetic electrons at the footpoints from the observed broken power-law photon spectrum ( assuming a low energy cutoff of $\epsilon_b$) \citep{2007ApJ...656.1187F, 1984ApJ...279..896N, 1971SoPh...18..489B}, which is shown as blue data points in Figure \ref{f4} \footnote{The energy deposition rate above $\epsilon_b$ should be considered as an upper limit of energy injection rate into the footpoints since the low cutoff energy of the injected electrons should be higher than $\epsilon_b$ \citep{2005A&A...435..743S}.}. The top panel for the different power components in Figure \ref{f4} shows that the phase of electron acceleration only covers the central period of a broad energy release period, which started at the onset of the flare and lasted for more than 2 minutes as indicated by the grey line for the observed total energy release rate including the energy increase rates of the {\it RHESSI} and {\it GOES} thermal components and the energy deposition rate by high energy electrons at the footpoints. The power carried by nonthermal electrons dominated for a period of less than 1 minute.

Although Figure \ref{f4} is similar to other flares studied before, there are several revealing details, which we discuss in the context of a stochastic accelerator operating in the coronal loop \citep{2009ApJ...702.1553L}. First, the appearance of the nonthermal component is associated with a suppression of the energy change rate of the {\it RHESSI} thermal component (see blue and red data points in the top panel of Figure \ref{f4}). This supports the stochastic particle acceleration model where nonthermal particles are accelerated from the high-energy tail of the thermal background. The acceleration may lead to an effective cooling of the background plasma as suggested by the {\it RHESSI} temperature evolution in Figure \ref{f1} due to efficient transport of high-energy electrons along the loop as revealed by the HXR footpoints. The decrease of the {\it RHESSI} temperature after the onset of electron acceleration is not an artifact of our spectral fitting model. By fitting the spectrum in the 6 to 15 keV energy range with an isothermal model, we obtain almost identical {\it RHESSI} temperature and $EM$. The appearance of radio emission produced by energetic electrons via the gyro-synchrotron process after the onset of electron acceleration also supports our spectral model (see discussion in the following subsection).
The {\it GOES} temperature increases monotonically in the rise phase and is not affected significantly by the onset of electron acceleration, which implies that the acceleration only proceeds in the relatively hotter {\it RHESSI} thermal component. The relatively cooler {\it GOES} thermal component likely originates from the evaporated chromosphere plasma and does not participate in the acceleration process directly.

Second, at the peak of $A$, the energy deposition rate by energetic electrons is comparable to the energy change rate of the {\it GOES} thermal component. However, the evolution of the {\it GOES} component is not well correlated with the nonthermal component. Given the short thermalization timescale at the chromosphere, this deposited energy must be radiated away at lower energy channels or drive evaporation of relatively cold plasmas \citep{1979ApJ...228..592B}. The total radiative energy of the {\it GOES} thermal component is about $10^{28}$ erg, which is a factor of a few lower than the energy content of the thermal component as shown in the middle panel of Figure \ref{f4} implying efficient emission at lower energy channels. Observations in the optical to EUV band will be able to distinguish these two scenarios \citep{2013ApJ...765...37F}. The middle panel of Figure \ref{f4} also shows that the total energy deposited by energetic electrons at the footpoints is also less than the energy content of the {\it GOES} thermal component, which is consistent with the lack of good correlation between the nonthermal energy deposition rate and the energy increase rate of the {\it GOES} thermal component.  These results do not support the theoretical Neupert effect which requires that the thermalization of the electron beam producing the observed HXR dominates the energy increase of the {\it GOES} thermal component \citep{1968ApJ...153L..59N, 2005ApJ...621..482V}.

 Last, the bottom panel of Figure \ref{f4} shows that the acceleration region of the {\it RHESSI} thermal component is expanding rapidly in the preheating phase presumably caused by the continuous heating. Although uncertainties in the {\it GOES} parameters in this phase render the quantitative detail questionable, the rapid increase of energy content of the {\it RHESSI} thermal component is evident.
 The expansion pauses at the onset of the electron acceleration possibly due to rapid energy loss as accelerated electrons leave the corona, and resumes before the disappearance of the broken power-law component.
 The conductive energy flux is given by $$F_{\rm cond}\sim 10^{26}(T_e/10^7{\rm K})^{7/2}(L/10^9{\rm cm}^{-1})(S/3\times 10^{16} {\rm cm}^2) {\rm erg\ s}^{-1},$$ where $L$ and $S$ are the loop length and cross section, respectively \citep{2006ApJ...638.1140J}. The increase rate of $U_{\rm GOES}$ is comparable to this rate and may be attributed to the thermal conduction. This conduction also leads to the rapid decrease of $U_{\rm RHESSI}$ and the associated volume decrease after about 02:18:30 UT. However there is little evidence to support a quantitative correlation between a static conductive heat flux from the {\it RHESSI} thermal component and the energy change rate of the other emission components, indicating that the evaporation flow or turbulence may modify the conductive heat flux significantly \citep{2006ApJ...638.1140J, 1978ApJ...220.1137A, 2009ApJ...702.1553L}.
 The energy decrease rate of $U_{\rm RHESSI}$ from 02:18:30 UT to 02:20:30 UT slightly exceeds the sum of $L_{\rm GOES}$ and the increase rate of $U_{\rm GOES}$, also suggesting radiation at lower energies.

\subsection{Electron Acceleration}

\begin{figure}[ht]
\begin{center}
\includegraphics[width=11cm]{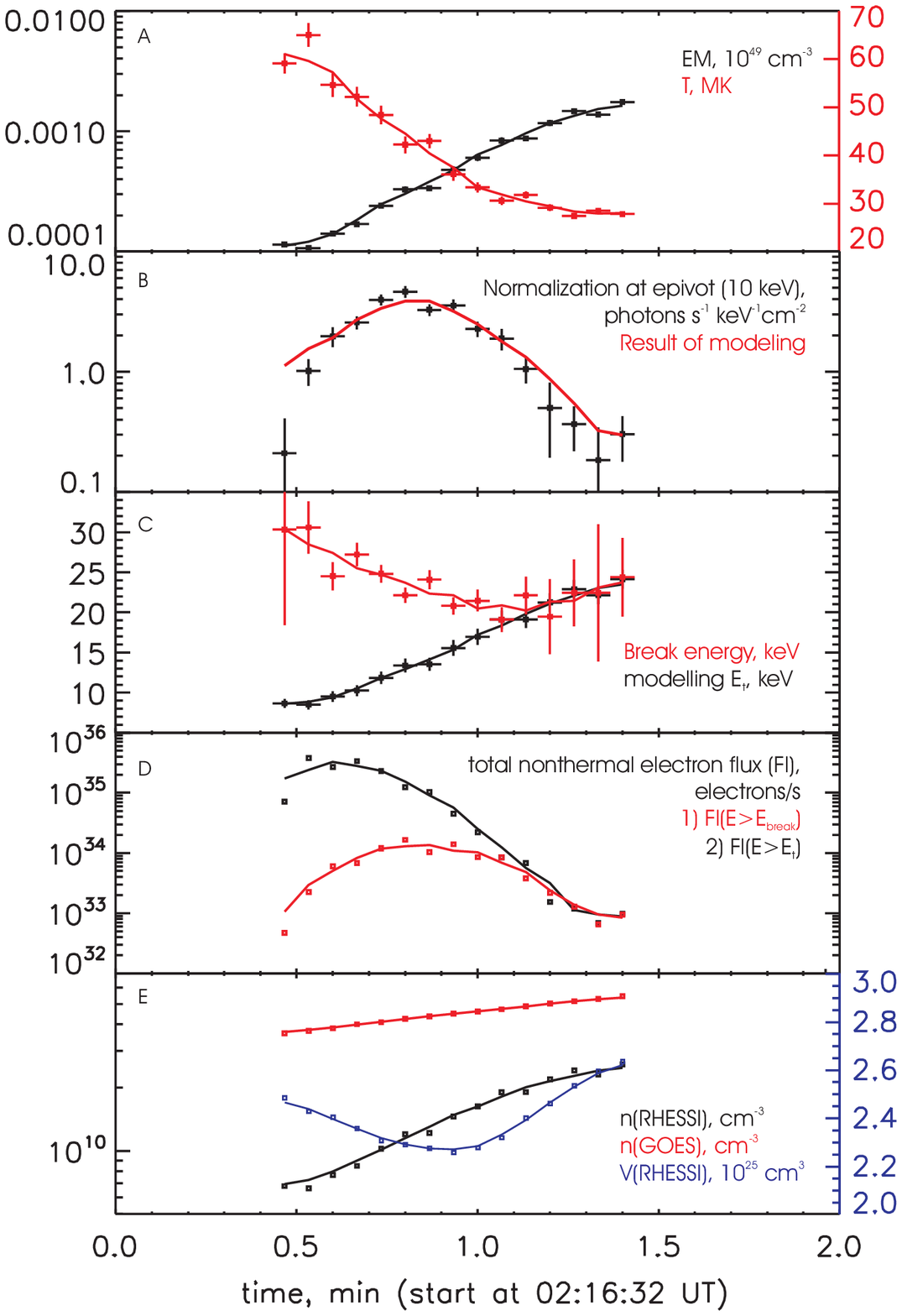}
\caption{
Spectral parameters for the model with $\gamma=4.0$ for the broken power-law component.
The first panel shows the temperature and $EM$. The second panel gives the normalization of the broken power-law component at the pivot energy of 10 keV. The solid line gives the result derived from other parameters with the stochastic particle acceleration theory. See text for details. The 3rd panel shows $\epsilon_b$ the break energy and $E_t$ the transition energy between the thermal and nonthermal component at the acceleration site. The 4th panel gives the flux of electrons injected into the footpoint 1) for $E> \epsilon_b$ and that escaping from the acceleration site 2) for $E> E_t$. The last panel shows the evolution of the density and volume of the {\it RHESSI} thermal component.
}
\label{f5}
\end{center}
\end{figure}

The good correlation between the radio fluxes and $A$ of the broken power-law component shown in panel 5 of Figure \ref{f1}, especially in the rise phase, strongly suggests that the radio and HXR emissions originate from the same parent electron population in the flare loop similar to those described in the `trap-plus-precipitation' scenario \citep{1976MNRAS.176...15M, 0004-637X-673-1-598}.
Panel 6 of Figure \ref{f1} shows that $\delta_H$, the spectral index between the 17 and 35 GHz bands,  and $\gamma$, the power-law photon index above the break energy, have the general soft-hard-soft spectral pattern, which further supports that the radio and high-energy HXR emission share a common high-energy electron population.
The fitted value of $\gamma$ is approximately constant at $\gamma\sim 4$. One can fix $\gamma=4$, which, in the context of stochastic particle acceleration, implies a constant level of turbulence \citep{2004ApJ...610..550P}. The reduced $\chi^2$ of a spectral fit with $\gamma=4$ is always less than 2.0 as shown in the last panel of Figure \ref{f1} \footnote{This $\gamma={\rm constant}$ assumption is different from what was proposed by \citet{2010ApJ...709...58L}.}.
To further explore the electron acceleration processes, we study this relatively simpler spectral model with $\gamma=4$ for the acceleration phase.

The model parameters are showed in Figure \ref{f5}, and we will explore the detailed characteristics of these nonthermal emissions in the context of a stochastic accelerator in the hot thermal component observed by the {\it RHESSI}. The electrons are then accelerated near the loop top. The radio emission is produced by high-energy electrons trapped in the loop, which have the same distribution as electrons in the loop top acceleration site. High energy electrons escaping from the loop will precipitate into the footpoints giving rise to the HXR emission there. In this model, the electrons producing the radio emission are related to the electrons producing the footpoints via an energy dependent escape timescale.

{\it Escape Timescale:}
The radio spectral index $\delta_H$ fluctuates around $\sim 1.5$ in agreement with the HXR spectral behavior. For gyrosynchrotron radiation, the corresponding index of the power-law electron distribution $\delta = (\delta_H+1.22)/0.9$ varies about 3.0 (see equation (35) given by \citet{1985ARA&A..23..169D}):
\begin{equation}
N = N_0 (E/\epsilon_b)^{-\delta},
\end{equation}
where $E$ is the electron energy and $N_0$ is a normalization factor for the energetic electrons.
According to the thick target model with $\gamma\sim 4$, the energy distribution of electrons injected into the target region of the footpoints has a spectral index of $\sim 5.0$ \citep{1971SoPh...18..489B}:
\begin{equation}
F = F_0 (E/\epsilon_b)^{-(\gamma+1)},
\label{flux}
\end{equation}
where $F_0$ is a normalization factor for the electron flux.
The escape timescale for electrons producing radio emission is then given by
 \begin{equation}
 T_{\rm esc}(E>\epsilon_b) \equiv {N \over F} = {N_0 \over F_0} \left({E\over \epsilon_b}\right)^{(\gamma+1-\delta)}\,.
 \label{esc}
\end{equation}
Using these expressions we see that for a major fraction of the acceleration phase, the escape timescale of electrons from the loop into the footpoints needs to have an energy dependence of $E^{\alpha}$ with $\alpha = \gamma-\delta_H/0.9-0.32/0.9$ varying about 2 \citep{0004-637X-673-1-598, 2010ApJ...709...58L}, which indicates more efficient trapping of higher energy electrons \footnote{These indexes have also been attributed to hardening of the electron distribution with the increase of energy with radio emission produced by higher energy electrons than HXR emission \citep{2011SSRv..159..225W}.}.

We note that $\delta_H$ should be considered as a lower limit of the spectral index of the power-law gyrosynchrotron emission since the source is not in the optically thin regime at 17 GHz. The values of $\alpha$ derived above should be considered as an upper limit, which is consistent with the positive values for $\alpha$ derived by \citet{2013arXiv1307.1837C}. Radio spectroscopic observations will be very helpful in this regard \citep{2008ApJ...677.1367A, 2011ApJ...731L..19F, 2013ApJ...768..190F}. The escaping particles are mostly associated with small pitch angles while the radio emission is dominated by electrons with large pitch angles. Polarimetric observations can be used to probe the pitch angle distribution of these emitting electrons.

{\it Acceleration Timescale:} According to the stochastic particle acceleration theory \citep{2004ApJ...610..550P}, nonthermal particles are accelerated from the thermal background plasma with the normalization of the nonthermal component determined by a transition energy $E_t$, at which the Coulomb collision timescale ($\propto E^{3/2}/n$) is equal to the acceleration timescale \citep{2010ApJ...709...58L}.
For an acceleration timescale $\tau_{\rm ac}\propto E^{-\beta}$, we then have
\begin{equation}
E_t = a n^{2/(3+2\beta)},
\end{equation}
where $n= (EM_{\rm RHESSI}/V_{\rm RHESSI})^{1/2}$ is the electron density at the acceleration site and $a$ is a parameter depending on the intensity of the turbulence responsible for the particle acceleration.
Then we have 
\begin{equation}
E_t = a (EM_{\rm RHESSI}/V_{\rm RHESSI})^{1/(3+2\beta)}\,,
 \label{et}
 \end{equation}
and, from the continuity of the electron distribution from low to high energies,
\begin{equation}
N_0 \simeq  {2\pi (EM_{\rm RHESSI}\ V_{\rm RHESSI} E_t)^{1/2}
 \over ( \pi k_{\rm B} T_e)^{3/2}} \exp{\left[-E_t\over k_{\rm B} T_e\right]}{\left(E_t\over \epsilon_b\right)}^{\delta}
\,,
\label{n0}
\end{equation}
where $T_e$ is the {\it RHESSI} electron temperature and we have ignored the effect of the power-law high-energy component on the electron density $n$ \citep{2010ApJ...709...58L}.

The normalization of the photon flux at $\epsilon_b$ is given by $A(\epsilon_b/10{\rm keV})^{-1.5}\epsilon_b^{\gamma}$, where $A$ is the normalization of the broken power-law component at the pivot energy of 10 keV. For a given turbulence intensity as suggested by the constant value of $\gamma$, the escape timescale remains the same. Then the normalization of the photon flux at $\epsilon_b$ is proportional to $N(\epsilon_b)\epsilon_b^{\delta}=N_0\epsilon_b^{\delta}$ \citep{1971SoPh...18..489B}, and we have the normalization of the broken power-law spectral component at 10 keV (See the 5th panel of Fig. 1):
\begin{eqnarray}
A &\propto& N_0 \left({\epsilon_b\over 10 {\rm keV}}\right)^{1.5+\delta-\gamma}
\nonumber \\
&\propto &  a^{1/2} EM_{\rm RHESSI}^{(2+\beta)/(3+2\beta)} V_{\rm RHESSI}^{(1+\beta)/(3+2\beta)} T_e^{-3/2}
\exp{\left[-a (EM_{\rm RHESSI}/V_{\rm RHESSI})^{1/(3+2\beta)}\over k_{\rm B} T_e\right]} \nonumber \\
&&\times {\left[a (EM_{\rm RHESSI}/V_{\rm RHESSI})^{1/(3+2\beta)}\over 10 {\rm keV}\right]}^{\delta}\left({\epsilon_b\over 10 {\rm keV}}\right)^{1.5-\gamma}\,,
\label{a0}
\end{eqnarray}
where we have used equations (\ref{et}) and (\ref{n0}) and $\delta \simeq 3$ in agreement with $\gamma=4$. Note that $\delta$ here is the spectral index for the electron distribution in the loop instead of that for the flux of electrons injected into the footpoints in the thick-target model. Equation (\ref{flux}) shows that the latter is given by $\gamma+1\simeq 5$.

For $\delta_H\approx 1.5$ and a roughly constant turbulence intensity as implied by $\gamma\approx 4$, $a$ is a constant, and one can use the time profiles of $EM_{\rm RHESSI}$, $V_{\rm RHESSI}$, $T_e$, and $\epsilon_b$ to reproduce the time profile of $A$ by adjusting $a$, $\beta$, and a normalization factor.
For the spectral model with $\gamma=4.0$ for the broken power-law component,
the first panel shows the temperature and $EM$ of the thermal component. The second panel gives $A$ the normalization of the broken power-law component at the pivot energy of 10 keV.
The parameter $\beta$ can take different values, depending on the acceleration models.
We choose $\beta=-0.2$, consistent with
 the stochastic acceleration timescale derived from {\it RHESSI} observations of a few flares \citep{2013arXiv1307.1837C}.
 For $a = 10 {\rm keV}/(10^{20}{\rm cm}^{-6})^{1/1.3}$ with $E_t\sim 13$ keV at the peak of $A$, the 2nd panel of Figure 5 shows a comparison between the $A$ derived from the spectral fit and that from equation (\ref{a0}).
The 3rd panel shows $\epsilon_b$ the break energy and $E_t$ the transition energy between the thermal and nonthermal component at the acceleration site. The 4th panel gives the flux of electrons injected into the footpoint 1) for $E> \epsilon_b$ and that escaping from the acceleration site 2) for $E> E_t$. The last panel shows the evolution of the density and volume of the {\it RHESSI} thermal component.


These results pose several challenges to the simple model proposed by \citet{2009ApJ...701L..34L}.
First, the 3rd panel of Figure 5 shows that $E_t$ is lower than $\epsilon_b$ in the first half of the acceleration phase and approaches $\epsilon_b$ later. \citet{2005A&A...435..743S} showed that the break energy of the observed photon spectrum should be lower than the cutoff energy of the injected electron distribution. Although the particle acceleration theory asserts that electrons are injected into the acceleration process above $E_t$ at the acceleration site of the {\it RHESSI} thermal component, only a fraction of these high-energy electrons with $E$ greater than $\epsilon_b$ manage to reach the footpoints producing the observed broken power-law HXR component. It is likely that the escape of high-energy electrons from the acceleration site of the {\it RHESSI} thermal component drives a return current. The corresponding electric field will confine lower energy electrons within the coronal loop \citep{1980ApJ...235.1055E, 2006ApJ...651..553Z, 2013ApJ...773..121C}.
Near the peak of HXR emission, the photon spectrum is about $ 1.5
(\epsilon/\epsilon_b)^{-4}$ photons cm$^{-2}$ keV$^{-1}$ s$^{-1}$, where $\epsilon$ is the photon energy. The corresponding spectrum of electrons injected into the footpoints is about $3.9\times 10^{33}(E/\epsilon_b)^{-5}$ electrons keV$^{-1}$ s$^{-1}$\citep{2007ApJ...656.1187F}, which gives an electron flux above $\epsilon_b\sim 22$ keV of $2.1\times 10^{34}$ electrons s$^{-1}$ (see the 4th panel).
The electron number density in the loop is about $10^{10}(EM/2\times 10^{45}{\rm cm}^{-3})^{1/2}(V/2\times10^{25} {\rm cm}^3)^{-1/2}$ cm$^{-3}$, where $V$ is the volume of the acceleration site (see the last panel of Figure 5). The total number of electrons in the acceleration site is about $2\times 10^{35}(EM/2\times 10^{45}{\rm cm}^{-3})^{1/2}(V/2\times 10^{25} {\rm cm}^3)^{1/2}$ leading to a specific acceleration rate above $\epsilon_b$ of $0.01-0.1$ s$^{-1}$ that is consistent with those obtained by \citet{2013ApJ...766...28G}.

One may extrapolate the flux of electrons injected into the footpoints with $E>\epsilon_b$ to $E_t<\epsilon_b$ to obtain the flux of electrons injected into the acceleration process at the acceleration site of the {\it RHESSI} thermal component. The 4th panel of Figure 5 also gives this electron flux above the transition energy $E_t$.
The corresponding specific acceleration rate can exceed $1$ s$^{-1}$, which is consistent with the Coulomb collision timescale of roughly $\sim 1.0 (n/10^{10}{\rm cm}^{-3})^{-1}(E/10{\rm keV})^{3/2}$ s. Therefore electrons in the {\it RHESSI} thermal component must be injected into the acceleration process very efficiently.
The high specific acceleration rate implies that the separation of the acceleration from the electron transport within the loop may be artificial. The acceleration and transport processes might be intimately coupled to give rise to the observed photon spectra. The spatial structure along the loop and the effect of the return current in the legs of the loop \citep{2006ApJ...651..553Z} need to be considered self-consistently to address this issue \citep{2009ApJ...702.1553L}.

Moreover, according to our assumption of a stochastic accelerator with $\beta=-0.2$ and the result shown in equation (\ref{esc}), the acceleration and escape timescales have different dependence on $E$, which according to the stochastic acceleration model will not lead to an exact power-law electron distribution \citep{2009ApJ...701L..34L}. Theoretically the distinction between thermal and nonthermal component is ambiguous in the acceleration region since acceleration, injection, and thermalization timescales are comparable near $E_t$. We expect a smooth electron distribution, which may be fitted with a thermal plus a power-law component \citep{2004ApJ...610..550P}. The stochastic acceleration theory we invoke here also assumes isotropic particle distributions, which may not be valid especially for high-energy electrons.
More detailed modeling is needed to address this issue.

\section{Conclusions}
\label{con}

 By combining in-depth data analysis with the theory of stochastic particle acceleration, we are able to explore the energy release processes in a simple flare loop that occurred on 12th August 2002 in unprecedented detail. We propose that the overall energy release in this flare may be divided into the following phases:
1) In the preheating phases, magnetic energy is released probably via some reconnection process and turbulence is generated at large scales via some instability. This large scale turbulence cascades toward smaller scales and at the same time energizes the bulk of background particles leading to heating and expansion of the energy release site;
2) In the rise phase of the nonthermal emission, the acceleration of electrons starts when the turbulence reaches small enough scales to interact with the high-energy tail of the thermal background via some resonant process and inject these electrons into the acceleration process. Since turbulence that is responsible for acceleration of high-energy electrons evolves, at large scales, over a long time, the system reaches a quasi-steady state with a nearly constant spectral index of accelerated electrons. The injection is very efficient at the beginning with the escape of high-energy electrons driving a return current and suppressing the escape of relatively lower energy electrons. Higher energy electrons can overcome the potential drop associated with the return current and produce HXR emission at the footpoints. The rapid energy loss caused by escape of high-energy electrons and their subsequent collisional loss in the footpoints also stops the expansion of the {\it RHESSI} thermal component. This period may be considered as the most dynamical phase with the electron acceleration, injection, thermalization, and transport processes strongly coupled. Multi-wavelength observations of this phase are highly warranted, which may reveal quantitative details of these physical processes \citep{2013ApJ...768..190F}.
3) In the decay phase of the nonthermal emission, the electron acceleration is suppressed by the increase of the density via chromospheric evaporation. Higher density gives shorter thermalization timescale leading to higher transit energy $E_t$ for given turbulence intensity; The {\it RHESSI} thermal plasma also resumes expansion in this decay phase. 4) The energy release rate is unable to balance the conductive energy loss rate and the {\it RHESSI} thermal component starts to cool down with the volume starting to shrink. The increase of the {\it GOES} thermal component appear to be driven mostly by conduction from the hotter {\it RHESSI} thermal component with no evidence for evaporation driven by higher energy nonthermal electrons whose energy content is also less than that of the {\it GOES} thermal component. The energy deposited at the footpoints by nonthermal electrons likely leads to heating of cooler and denser plasmas deeper in the chromosphere, which may give rise to radiation in the optical to EUV range. 5) After the overall energy release ceases, the energy increase of the {\it GOES} thermal component and the related radiative energy loss is less than the energy loss of the {\it RHESSI} thermal component via conduction, also indicating efficient emission at lower energies.

Although these results are consistent with the stochastic particle acceleration theory in general, they also suggest processes, e.g. return currents in the early acceleration phase, not covered in the simplest stochastic particle acceleration model \citep{2009ApJ...702.1553L}. This study also shows that to model a specific flare in detail, the evolution of the source structure must be considered. The process of electron acceleration is intimately coupled to the electron transport processes and small scale turbulent plasma waves play an essential role in the onset of electron acceleration within a more gradual energy release process. More quantitative modeling, detailed polarimetric and/or spectroscopic observations in radio and X-ray bands, and simultaneous optical to EUV observations are warranted to advance our understanding of the related physical processes.



\acknowledgements
Dr. S. Liu and I. Sharykin thank Drs. W. Q. Gan from the PMO and A. Struminsky from the IKI (Moscow) for making this collaborative work possible. This work is partially supported by the Strategic Priority Research Program - The Emergence of Cosmological Structures of the Chinese Academy of Sciences, Grant No. XDB09000000, the NSFC grants 11173064, 11233001, and 11233008, project RFBR No13-02-91165, and by the EC's SOLAIRE
Research and Training Network at the University of Glasgow
(MTRN-CT-2006-035484), by Rolling Grant ST/I001808 from the UK's
Science and Technology Facilities Council, and by the EC's HESPE project (FP7-SPACE-2010-263086).

\end{document}